\documentclass[11pt,twoside]{article}

\usepackage{asp2006}

\begin{document}
\newcommand{\rcite}[2]{{\bf #1.~#2}}
\title{Extragalactic Jets - Reflections on the Conference}   %%% Fill in title
\author{Lawrence Rudnick}   %%% Fill in author names
\affil{Department of Astronomy, University of Minnesota}    %%% Fill in author affiliations

\begin{abstract} %%% Abstract to run on from here.
 I review some of the important and exciting recent advances that were presented at the 2007 conference on Extragalactic Jets in Girdwood, Alaska, using as a framework the scientific challenges presented by R. Blandford at the beginning of the meeting. Sprinkled throughout are thoughts about the marvelous prospects for jets in the next several years, as a host of new observatories mature and simulations reach new levels of sophistication.
\end{abstract}

\subsection*{Perspectives}
On the one hand, the progress we have made in characterizing the properties and understanding the physics of jets is simply breathtaking -- from the rapidly maturing fields of TeV observations and mm VLBI, to the heroic surveys of enormous samples to examine statistical trends, to the still accelerating ability to incorporate physics and probe multiple scales in MHD, even relativistic, simulations.  On the other hand, the big questions we are asking today  - such as the launching and the content of jets are depressingly similar to those we asked ten and even twenty-five years ago \citep{copp97,per83}.  Nonetheless, the value of astrophysical jet studies appears to grow with time;  this was documented by my recent search on Amazon, where we find {\it Astrophysical Jets} from a 1992 conference selling for \$100.00 (US), while {\it Astrophysical Jets} from a 1982 conference goes for \$288.00 (US)!

These reflections are meant to be read in conjunction with the challenges provided by \rcite{R}{Blandford} at the beginning of the meeting.  I highlight a few of the interesting new findings presented at the meeting in each of these areas (liberally interpreted),  along with some personal thoughts about the needs and opportunities for progress in the next several years. References are in general limited to these proceedings, indicated by the name of the presenter in {\bf boldface}.
The electronic contributions can be found at\\ http://aftar.uaa.alaska.edu/jets2007/program.html.

\subsection*{Challenge 1:  Locate the sites of radio, $\gamma$-ray emission}
The increasingly detailed information coming from HESS has created new challenges for understanding the sites of pc-scale emission.  \rcite{S}{Wagner} presented variability data for energies above 200 GeV  in PKS 2155-304, showing up to 100\%  variations on time scales of minutes. This creates a dilemma because if the variable regions come from far out in the jet, the area corresponding to the variable region should only be a tiny fraction of the inferred cross section.  However, if the emission arises much closer in, then the much higher  opacity should prevent the $\gamma$-rays from emerging. On a brighter note,  \rcite{A}{Levinson} showed how GLAST observations can provide sophisticated tests of jet models -- e.g.,  the changes in $\gamma$-ray opacity (leading to pair-production) as a function of distance along the jet would produce flares that propagate from low to high energies.

 At mm wavelengths, \rcite{A}{Marscher} cautioned us,  compact base regions of jets may be ``pseudo-cores'' at $\sim$pc from the black hole; this is evidenced in 3C120  by the delay between dips in the X-ray brightness and the onset of superluminal motions 60 days later. Further out, over 100 pc from the nucleus of M87, jet knot HST-1 shows not only multi-band variability, but seems to be the origin of its own superluminal ejections, at speeds of 1-4c (\rcite{D}{Harris}, \rcite{T}{Cheung}).  Harris cautions that we should not think of this as an actual ejection site, but a slight local disruption in the flow that renders it radiative at that point.  It will be interesting to see if we can find confirming evidence for such an obstruction(ist) view of jets.

On pc scales, we are now beginning to resolve jet transverse structures; observations of limb brightening and polarization changes (\rcite{R}{Sambruna}, \rcite{D}{Gabuzda}) are sometimes interpreted as fast spine, slow sheath structures. On kpc scales, where the relativistic effects are mild at best, transverse velocity gradients provide a better description of the data (\rcite{R}{Laing}), with a combination of projected toroidal and longitudinal fields.  A great deal has been learned from the jet/counterjet comparisons on these larger jets, and such comparisons on pc scales, while difficult, will be essential for separating out both the dynamics and sites of radio emission.

Results from the gamma ray observatories may hold some real surprises.  Correcting for the extragalactic background is still a tricky business, but \rcite{A}{Wolter} (using INTEGRAL) and \rcite{S}{Wagner} (using HESS) reported results on z$>$0.1 blazars that showed still rising spectra beyond 100 keV and $\approx$TeV (!), respectively.  Wagner's reported data on 1ES1101-232 showed that the lower energy spectrum had already turned over in the X-ray regime, so the gamma rays are most likely the Inverse Compton boosted 10$^{14}$ Hz photons, requiring $\gamma > 10 ^5$.

\subsection*{Challenge 2: Map the velocity fields}

\rcite{Y}{Kovalev} showed a spectacular $104$:1 dynamic range 15~GHz image of the M87 jet and counterjet, complemented by \rcite{C}{Walker's} exciting early movie made at 43~GHz using the VLBA.  The potential for measuring both jet and counterjet motions will greatly enhance our ability to model these relativistic motions.  And none too soon, because as \rcite{D}{Hough} showed, things are getting messy on scales of 10s of pc, where {\it accelerations} are becoming common;  \rcite{T}{Krichbaum} cautioned, however, that beaming effects from a fast spine/slow sheath structure (or transverse velocity gradients in general) can confuse the issue.   Non-ballistic, even transverse  motions were shown in NRAO~150 by \rcite{I}{Agudo}, and in 1253-055 by \rcite{M}{Lister}.   When you're sure you can spare a couple of hours, go to their MOJAVE website, http://www.physics.purdue.edu/MOJAVE, and watch the movies.

On kpc scales, \rcite{R}{Laing} and his collaborators continue their lone {\it tour de force} reconstruction of velocity and other parameters through the flaring regions from FRI jets such as 3C31 and NGC~315.   They find drops in velocity from $\approx$0.9c to $\approx$0.3c through this region, and smooth transverse velocity gradients when the jets becomes resolved.   It's time for the simulators to catch up with these beautiful data.

One issue not currently well-studied is the connection between flows on pc and kpc scales.  I thought the magnificent work on 3C120 twenty years ago \cite{walker} would set the stage for many more such investigations, but it didn't.  At this meeting, \rcite{G}{Giovannini} presented work on MK501 showing very strong changes in angle between small and large scales, and \rcite{I}{Agudo} showed evidence for alignment changes $>$120$^o$ in NRAO~150. We need more such examples, along with better tools to disentangle beaming effects from large intrinsic angle variations.

Compact symmetric objects (CSOs) remain a hot topic (contributions from {\bf M. Lister, M. Giroletti, L. Stawarz, M. Orienti, H. Nagai}), and allow a nice measurement of velocity of the (assumed) jet terminus, and thus a kinematic age for the source.  There continue to be too many of these 10$^3$ year old sources if the 100~kpc sources are their long-lived progeny.  Whether they turn off, or are stifled by ISM interactions (see more below) is still unclear. \rcite{R}{Morganti} reported that large HI disks are found in the host galaxies of compact (10kpc) radio sources only  -  not in those of large radio sources.  This is a very important clue to the evolution of CSOs, although it is not yet clear whether the HI disks indicate a jet-frustrating barrier or the residual of a turned-off fuel supply.

\subsection*{Challenge 3: Identify the emission mechanism}

In jets, the radio emission comes from synchrotron emission --  it feels so good just to say something we do know!  But the higher energy story is still unclear.  \rcite{M}{Hardcastle} described how in the jet of Centaurus~A,  the X-ray knots always have radio counterparts but not vice versa, and X-ray bright knots are stationary while radio features can move at speeds up to 0.5c .  In the 3C33 hotspot, neither synchrotron nor external Inverse Compton (IC) models are successful at connecting the radio and X-ray emission.   He, \rcite{M}{Erlund} and others showed offsets between  X-ray and radio hotspots, so we're missing some key piece of the puzzle here. Even when the radio and X-ray components appear coincident, as in the PKS1127-145 jet presented by \rcite{A}{Siemiginowska}, one component models appear insufficient.  Can we get any insights from observations such as presented by \rcite{D}{Schwartz} on PKS~1055+201, where the high frequency radio jet is surrounded by a broader region seen in both (IC) X-rays and low frequency radio emission?

In a completely different regime, the longstanding questions about $>$100~GeV emission mechanisms -  $\pi_0$ decays from hadronic interactions or Inverse Compton emission from the optical synchrotron electrons - remain.  The situation is improving quickly from the observational side, as seen in the multi-frequency variability presentations of \rcite{D}{Paneque} and \rcite{K}{Lee}, and optimists can hope for a speedy resolution.  \rcite{R}{Wagner} presented some spectacular MAGIC variability date on two flares in MK~501 with variations on time scales $<$3~minutes! This is a unique (but hopefully just the beginning) set of observations, showing increasing variability and a shift in peak time with higher energy. He argued that such variations are unlikely to arise in hadronic models, although the lack of correlation with optical or X-ray variability means the standard Inverse Compton models are also problematic. Good quality, simultaneous data are sorely needed.

\subsection*{Challenge 4: Understand the changing composition}

\rcite{G}{Madejski} presented some tantalizing evidence from SUZAKU and SWIFT for ``bulk Compton" emission in PKS1510-089.  This effect, predicted twenty years ago by \cite{beg87}, relies on Compton scattering of the accretion generated photons by the overall jet flow, rather than from shock structures in the jet. It implies that the protons are dynamically dominant in the jet, as does the \rcite{T}{Cheung} analysis of the broadband emission from the Cygnus~A hot spots.

On kpc scales, where we have lots of observational data, this issue is still quite murky, with few ways to distinguish between e$^+$e$^-$ and p$^+$e$^-$ plasmas. \rcite{R}{Laing} reports the emergence of a characteristic density, $\rho$=1.4 - 2.4$\times$10$^{-27}$ kg/m$^3$ based on analysis of 3C296, 0326+396 and 3C31 (although why he insists on S.I. units is a great mystery). Despite this exquisite measurement, he can still not rule out either possibility, although an e$^+$e$^-$ jet would require the injection of some thermal material or some otherwise unobserved excess momentum component at low energies.

Down near the black hole, Poynting flux driven jets have long been the darling of theorists, but have never captured the hearts of observers working on larger scales.  However, 3D MHD simulations of electromagnetic jets may now help them become serious competitors, due to the work of \rcite{M}{Nakamura}, \rcite{H}{Li} and collaborators.  \rcite{H}{Li} showed that the magnetic domination could extend to the scales of radio galaxy lobes, and provide a serious alternative to particle-dominated models.  Interactions with the external environment would be different for such plasmas, and this intriguing possibility should provide welcome competition for our standard pictures.

\subsection*{Challenge 5: Measure external pressures  (ISM interactions)}
Here I concentrate on the variety of smaller scale ISM interactions -- this topic, long on promise for decades, is really coming of age!  These interactions certainly have pressure signatures in them, but the dynamics both confuse and enrich the issue.  \rcite{H}{Nagai} presented a summary of CSO component velocities -- their random directions argue for strong environmental interactions.   \rcite{M}{Lister}, e.g., showed a gorgeous image of the Seyfert galaxy NGC 4151 (from \cite{mundell}), illustrating the relationship of the jet to the molecular hydrogen torus, the HI ring, and the central ionized region.  Similarly, the AGN jet in NGC~4258 is exciting the surrounding dust and gas, as seen from Spitzer, Chandra and VLA observations \citep{yang07}. \rcite{P}{Ogle} reported the common presence of high ionization lines from silicates around AGN, with 1000 year cooling times that argue for continuous reheating. \rcite{R}{Morganti} detects broad HI absorption in addition to the previously known narrow absorption; in a number of cases, these broad lines and high, blue-shifted velocities  provide evidence for  AGN-driven outflows, with energy transfers comparable to starburst-driven superwinds.

\rcite{P}{Ogle} also shocked us with the report that shocked H$_2$ emission was common and strong around radio galaxies, sometimes representing $\approx$15$\%$ of the total IR luminosity.  All of the detected systems have morphological peculiarities, so the effects of the jet are widespread.  He suggested that we might more properly call these systems MOHEGs (MOlecular Hydrogen Emission Galaxies) instead of radio galaxies;  such sacrilege must be snipped in the bud!

\subsection*{Challenge 6:  Deduce jet confinement mechanism}
This topic is closely tied to the structure of the magnetic field in jets, approached through VLBI polarimetry, rotation measure and other studies (see contributions by \rcite{D}{Gabuzda} (and her legions), {\bf P. Veres, S. O'Sullivan, T. Savolainen, M. Mahmud, K. Asada \& M. Inoue}).  These efforts are bringing the VLBI polarization studies to an almost mature level, where one can begin to assess statistical samples -  critical to separate out the weather.  Gabuzda emphasized, e.g.,  a study showing the excellent consistency between measures of helicity from Faraday rotation gradients and from the circular polarization Faraday conversion. This field (no pun intended) is not for the faint of heart, but critical for understand field geometries, and ultimately, the possibilities for magnetic confinement and relativistic particle acceleration.  The simulations of Poynting flux driven jets discussed earlier are also critical for these confinement and acceleration issues. In particular,  we need to understand how to take such jets from the SMBH out to pc scales, and also get predictions about the relationships between pc and kpc scale properties. Modelers of all religions should prognosticate in anticipation of VSOP-2 in 2012, rather than waiting for an {\it ex post facto} epiphany.

The connection between hot spot structure and jet properties  - clearly related to jet confinement on large scales - has not received the attention it deserves.  As part of her attempt to derive cosmological tests with radio galaxies, R. Daly suggests that larger beam powers lead to more complex hot-spot structures (side-to-side comparisons).

\subsection*{Challenge 7: Infer jet powers, thrusts}

These are exciting times for measuring jet powers, because of the evacuation of X-ray cavities in clusters, as discussed in more detail below.  Detailed dynamical models of FRI sources, as presented by \rcite{R}{Laing}, provide a complementary way to calculate momentum and energy transfers, and he showed the initial steps in bringing these two types of estimates together.  Regarding jets in FRII (classical double) sources, \rcite{R}{Daly} derived beam powers of 10$^{45-6}$ erg/s, while \rcite{M}{Hardcastle} admonished us to be aware of the complex interactions between the jet and its surrounding cocoon, different from the situation for naked FRI jets.

\subsection*{Challenge 8: Test and apply the `central hypothesis'}

One key way to examine the black hole paradigm (mass, spin, accretion rate  plus orientation $\longrightarrow$ observations) is through statistical properties of large samples.  It was heartening to see so much work in this area, both in making extensive new observations (e.g., Chandra jets - \rcite{J}{Gelbord}; VLBI mapping - \rcite{G}{Taylor}; MAGIC VHE $\gamma$s - \rcite{R}{Wagner}), and in developing new ways of looking at the data ({\bf I. Fernini, G. Fossati, L. Stawarz}).  \rcite{D}{Evans}, e.g., using Chandra observations of 3C radio galaxies suggested a new classification into two classes based on the efficiency of their accretion rates.  \rcite{L}{Stawarz} separated AGNs into two sequences on the radio-loudness, Eddington-ratio plane, corresponding to the elliptical or disk nature of the host galaxy.

Lively and healthy discussions accompanied all such attempts to unify.  \rcite{E}{Valtoaja} gave a comprehensive review of current studies correlating AGN properties with the underlying physical quantities. Although he was somewhat discouraged about our progress (perhaps it's the long nights above 60$^o$?), my feeling is that the admittedly chaotic approach to unification studies is exactly what is called for now -- we are scrambling to find the right observational keys and avoid the pitfalls of orientation biases. Looking forward, \rcite{G}{Madejski} highlighted the progress expected from GLAST and Suzaku studies, with time resolved spectra during outbursts and the ability to probe relativistic particle energy ranges (through Inverse Compton radiation) that are inaccessible in the radio.  Overall, the central hypothesis seems to be intact, but validating it is still a major challenge.

\subsection*{Challenge 9: BHGRMHD Capability}
[{\it Black hole, general relativistic magnetohydrodynamics - well, not yet, but some progress!}]  One line of promising work is on the analytical side, such as the analysis of helical modes and instability growth by \rcite{M}{Perucho}.  \rcite{Y}{Mizuno} presented very ambitious simulations of a jet spine driven by magnetic fields threading the ergosphere, with a broader sheath wind driven by magnetic fields anchored in the accretion disk;  2D results suggest that such jets can be stabilized against Kelvin-Helmholtz instabilities.

Although the above work is critical, we are really negligent in investigating how physical properties of jets on pc-scales connect with observational data - even very basic issues that we've known for years such as distinguishing between bulk and pattern speeds.  Some effort is going into this, but much more is needed.  \rcite{P}{Wiita} described an analytical investigation of jet opening angle and velocity gradients, and how these would be manifest in VLBI observations.  This yields, for example, an explanation for the relatively few highly superluminal components in TeV blazars.  \rcite{J}{Marti} presented a very nice set of simulations, which in the end I found discouraging from the standpoint of diagnostics, showing that classical and relativistic jet models yielded virtually equivalent results once scaled by the internal beam Mach number.

\rcite{C}{Swift} presented a relativistic ray-tracing program to actually create pseudo-observations from simulations of relativistic flows.  In this way she can ``reverse-engineer" bright spots in a jet flow, and has already found, e.g., that retarded time effects can have a major influence on the appearance of specific features. Why should we sit around guessing at what we're looking at? More work like this must be done!

\rcite{T}{Krichbaum} emphasized the potential mm-VLBI to reach to the regime of a few Schwarzschild radii, if the world's large mm facilities join in the global network.  This is an opportunity not to be missed. Already, the base of the M87 jet can be limited to $<$15 R$_S$, which suggests jet launching mechanisms closely related to the SMBH.

\subsection*{Challenge 10: Quantify the role of jets in clusters of galaxies}

We are in a whole new era in jet/cluster physics, nicely summarized by \rcite{B}{McNamara}. With the jet's environment actually visible in X-rays, and likely more uniform than the ISM (although this bears scrutiny), there is an enormous potential to derive basic physical quantities about jets such as their pressures and total deposited energy.  Added to this, we have a whole new community interested in what influence jets have on cluster plasmas, since the X-ray properties present puzzles such as how to shut off the cooling in the cores. \rcite{J}{Croston} showed us the very pretty result that clusters with radio galaxies are hotter, at a given X-ray luminosity, than those without current radio galaxies, arguing for an AGN heating role.

\rcite{B}{McNamara} reminded us that in most cluster cavities excavated by jets the X-ray rims are cool, not shocked.  This points to (trans)sonic processes, but often with enormous energy inputs (up to 10$^{62}$ ergs in the $>$200 kpc cavity around MS~0735.6+7421, where a mild M$\sim$1.3 shock is seen).  Another important lesson learned from cluster jets is that there is an enormous range in radiative efficiency, i.e., you can't use the radio power to estimate the underlying jet power.

A number of other lines of inquiry are helping us understand jet/environment interactions, including the intermittent jet investigations presented by \rcite{M}{Jamrozy} and the exquisitely detailed rotation measure analysis of FRIs by \rcite{R}{Laing}. Inverse Compton emission in lobes (\rcite{J}{Goodger} and \rcite{G}{Migliori}) show that there are variations in the ratio of magnetic to relativistic particle pressures that need to be sorted out before we can understand the environmental role. \rcite{J}{Croston} showed us the shielding power of cocoons;  when jets are ``naked" in a cluster, they need about 10$\times$ more pressure to match the external pressures - is entrainment the key?  \rcite{V}{Gaibler} explored the importance of underdense jets, whose weak shocks may provide a quite effective thermalization mechanism, driving cluster bubbles and heating.  Another perspective on these interactions will come from further X-ray observations of the 3C186 cluster, as reported by \rcite{A}{Siemiginowska}, since this (possibly re-starting?) compact radio source is likely to be only 10$^5$ years old.

 We are still in the dark regarding AGN fractions, duty cycles and lifetimes of cluster galaxies, especially given the warning about radiative efficiency.  But the striking deep Westerbork maps of Abell 2255 by \rcite{R}{Pizzo} admonish us that an entire cluster could be stirred up and heated by many tailed radio galaxies.  Is this really a physically exceptional system, or just a statistical fluke regarding the number of AGN active at once? And will the shallower groups of galaxies, where the gas is more easily disturbed, the next frontier for jet environment interactions, as suggested by \rcite{E}{Freeland} and \rcite{E}{Wilcots}?

Complementing the enormous progress in jet/cluster studies, a new generation of simulations is beginning to improve both our diagnostic abilities and address the issues of cluster heating. \rcite{T}{Jones} presented a superluminal tour through the relevant MHD physics.  \rcite{S}{Heinz} showed quite promising 3D simulations of turned-off jets in a dynamic ICM, where the pressure-driven wakes generated quite isotropic large-scale turbulence. Jet heating thus has the potential for meeting both the energetics and isotropy needs to quench cooling.  At the next level of physical sophistication, \rcite{D} {DeYoung} presented a more cautionary tale from studies of 3D MHD simulations of jet/cluster interactions.  He emphasized that the 3D MHD is essential for generating mixing and lifting of cluster plasmas, while the magnetic fields suppress instabilities -- leading to longer mixing timescales.  \rcite{T}{Jones} presented a closer examination of the magnetic field evolution, showing that (only in 3D) vortices form which greatly amplify and self-organize the fields to where they can dominate the flow. With these and related efforts, the heating issue for clusters seems within our grasp.

\subsection*{Extra Challenge: Characterize Relativistic Particle Acceleration}

Although this fell on the cutting room floor of \rcite{R}{Blandford}'s  original challenges, it remains a critical issue on which he has also done seminal work.  Blandford reminded us of one key observational puzzle, how the inter-knot acceleration takes place in M87's jet; second order processes are probably needed.  \rcite{D}{Schwartz} argued that the same challenge holds for PKS~1055+201, in addition to whatever happens at the jet shocks. \rcite{R}{Perlman} reported on the optical polarimetry of FRI jets, with no clear pattern in the relationships between the polarized radio and optical and the X-ray maxima; he argues that particle acceleration must thus proceed via more than one mechanism in these jets.

At the low energy end, \rcite{R}{Laing} presented evidence for a robust (against local conditions) low frequency relativistic electron slope of -2.2 to -2.1 (see also \cite{young}), creating a puzzle for how particle acceleration is so regulated.  Multiwavelength data are critical and becoming more common, such as those presented by \rcite{T}{Cheung}, using Spitzer data, and \rcite{S}{Jorstad}, combining radio, IR and X-ray data on 1317+520, arguing for relativistic electrons with energies up to 100 TeV!

\rcite{K}{Blundell} reminded us of the importance of tracking all of the relativistic plasma (and not simply relying on the currently observed luminosity), by putting in appropriate physics with respect to jet intermittency and rapid fading of radio lobes.  Although her presentation was aimed at investigations using radio galaxies as large scale structure diagnostics, it is also important for understanding the lifetime history of particle acceleration.

\rcite{R}{Protheroe} deserves kudos for an extremely innovative proposal to use fossil AGN as the source of ultra-high-energy cosmic rays.  The basic idea is that fossil jets self-organize into a stable reverse-field pinch (RFP) configuration. The slow decay of this field through reconnection induces electric fields of order 10$^{-5}$V/m, which can then accelerate seed cosmic rays to high energies. Whether or not this RFP mechanism will ultimately work, it may provide the stimulus for more creative thinking about the sources of the highest energy cosmic rays.

Relativistic particle acceleration seems ripe for progress in the coming years, with our increasing ability to probe the low energy end of the electron distribution (through Inverse Compton emission against various backgrounds) and the high electron energy end with X-rays, as well as the enormous promise of UHE cosmic rays.  Add to this the increasing attention to rapidly varying sources, including in $\gamma$-rays, and maturing numerical simulations, and particle acceleration may yet escape its current low profile!

\subsection*{Take-Home Messages}

As an observer, I see us about to enter a new golden age of extragalactic jet studies, pushing the limits of resolution and photon energy into untapped regimes.   Newly maturing facilities span Earth and space, including the High Energy Stereoscopic System (HESS), GLAST  and MAGIC on the high energy end, the Long Wavelength Array (LWA) and the extended Low Frequency Array (eLOFAR) at the low end, and -- in the barely explored millimeter regime -- the Atacama Large Mm/sub-mm Array (ALMA) and mm VLBI, and, for  unparalleled resolution,  VSOP-2 and SIM.  And if the funding deities remain beneficent, we may also see a great deal more of what the VLBA can do. Theoretical tools are similarly advancing as computer power expands through, e.g.,  multi-processor systems, but the physics problems are hard, and more bright people are needed. So stay tuned!

Following are short take-home messages provided, with and without attribution, by courageous meeting participants:

\begin{itemize}

\item{We observers need accurate and believable jet power estimates to assess the importance of AGN feedback on galaxy evolution! [P. Ogle]}
\item{Save the VLBA!}
\item{What we see (ultra-relativistic electrons + B [synchrotron] + U($\gamma$) [Inverse Compton] ) is NOT the medium that transports energy on Mpc scales.  Therefore, the assumption that proper motion of emitting blobs tells us jet velocity is dubious at best. [D. Harris]}
\item{Expanding VLBI to shorter wavelengths ($<$1mm) would allow us to image the jet base of AGN with $\approx$20~$\mu$arcsec resolution.  This corresponds to a few R$_s$ and will allow us to better understand jet launching.  The first step is done:  M87 has been imaged with $\leq$15$\times$ R$_s$ at 3mm. [T. Krichbaum]}
\item{I think it would be possible to image the accretion disks in AGN by VSOP-2. [H. Nagai]}
\item{The study of jets covering a wide range in orientation is difficult, but will remain critical to test jet models in the future.}
\item{Support high resolution imaging and timing studies of jets, and save the VLBA! [A. Marscher, S. Jorstad et al.]}
\item{Any statement about magnetic fields in radio jets should be supported by a calculation.}
\item{Will GRMHD simulators please decide on a couple of test problems so that codes can be compared sensibly?  Only then can we start to believe the results. [P. Wiita]}
\item{It will be extremely important if we can convince ourselves that extragalactic jets really do have helical B fields -- keep this possibility in mind when analysing your results! [D. Gabuzda]}
\item{Helical B field, why not? [K. Asada]}
\item{MHD driven mechanisms are still promising rather than other process in extragalactic jets, but
we need to more thoroughly explore the transition to larger scales. [M. Nakamura]}
\end{itemize}

\subsection*{Quotables}
Finally, I close with some choice quotations from the conference that hopefully convey a little of the fun that we all had.   I give the only the initials of the speakers, to provide a patina of confidentiality.

\begin{itemize}
\item{``When you see something at 4c, you have to deal with it" (DH)}
\item{``How can you observe Cherenkov radiation while the Moon is out?"...\\
  ``To my mind, it's a philosophical issue" (RW)}
\item{``Despite the fact that I'm one of your collaborators, Dan ...." (EP)}
\item{``When is a knot not a knot?" (MJH)}
\item{``But you can't violate causality..."  (AM)\\  ``I think we should keep an open mind..." (DG)}
\item{``I don't believe in particles anyway." (DH)}
\item{``I don't care what the answer is;  I don't understand it in any case." (LR)}
\end{itemize}

\acknowledgements %%% Text of acknowledgements runs on after this command.
Many thanks to Travis Rector and Dave DeYoung for organizing such a great conference, and enabling me to pay the closest attention I ever have to four days of exciting talks.  Thanks to lots of people who corrected mistakes in these reflections;  readers are still cautioned not to cite this work as a source of factual information! Instead, go directly to the meeting contributions when they appear on astro-ph or in the ASP Conference Proceedings.  My participation at the conference and my AGN-related research is supported at the University of Minnesota through grant AST~06-07674 from the U.S. National Science Foundation.

\end{document}